\documentstyle[times,pramana,epsf,epsfig,floats,amsmath,axodraw,color]{ias}

\input paperdef

\begin{document}



\thispagestyle{empty}
\setcounter{page}{0}
\def\thefootnote{\fnsymbol{footnote}}

\begin{flushright}
hep-ph/0611374
\end{flushright}

\vspace{1cm}

\begin{center}

{\large\sc {\bf The Road Towards the ILC: Higgs, Top/QCD, Loops}}
\footnote{plenary talk given at the {\em LCWS06}, 9-13 March
  2006, Bangalore, India}

\vspace{1cm}

{\sc S.~Heinemeyer$^{1\,}$%
\footnote{
email: Sven.Heinemeyer@cern.ch
}%
}

\vspace*{1cm}

$^1$ Instituto de Fisica de Cantabria (CSIC--UC), Santander,  Spain 

\end{center}

\vspace*{0.2cm}

\BC {\bf Abstract} \EC
The International Linear $e^+e^-$ Collider (ILC) could go into
operation in the second half of the upcoming decade. Experimental
analyses and theory calculations for the physics at the ILC are
currently performed. We review recent progress, as presented at the
{\em LCWS06} in Bangalore, India, in the fields of Higgs boson physics and
top/QCD. Also the area of loop calculations, necessary to achieve the
required theory precision, is included.

\def\thefootnote{\arabic{footnote}}
\setcounter{footnote}{0}

\newpage


\title{The Road Towards the ILC: Higgs, Top/QCD, Loops}

\author{S. Heinemeyer$^1$}
\address{$^1$Instituto de Fisica de Cantabria (CSIC-UC), Santander, Spain}


\pacs{2.0}

\abstract{
The International Linear $e^+e^-$ Collider (ILC) could go into
operation in the second half of the upcoming decade. Experimental
analyses and theory calculations for the physics at the ILC are
currently performed. We review recent progress, as presented at the
{\em LCWS06} in Bangalore, India, in the fields of Higgs boson physics and
top/QCD. Also the area of loop calculations, necessary to achieve the
required theory precision, is included.

}

\maketitle

\definecolor{Blue}{named}{Blue}
\definecolor{Lblue}{cmyk}{1,0,0,0}
\definecolor{Red}{named}{Red}
\definecolor{Green}{named}{PineGreen}
\definecolor{Black}{named}{Black}
\definecolor{Magenta}{named}{Magenta}
\definecolor{Royal}{named}{RoyalBlue}
\definecolor{Orange}{named}{Orange}
\definecolor{Purple}{named}{Purple}
\definecolor{YOrange}{named}{YellowOrange}
\definecolor{Yellow}{named}{Yellow}
\definecolor{Mahogany}{named}{Mahogany}
\definecolor{Brown}{named}{Brown}
\definecolor{Gray}{named}{Gray}
\newcommand{\black}[1]{\color{Black}#1 \color{Black}}
\newcommand{\gray}[1]{\color{Gray}#1 \color{Black}}
\newcommand{\red}[1]{\color{Red}#1 \color{Black}}
\newcommand{\blue}[1]{\color{Blue}#1 \color{Black}}
\newcommand{\lblue}[1]{\color{Lblue}#1 \color{Black}}
\newcommand{\green}[1]{\color{Green}#1 \color{Black}}
\newcommand{\magenta}[1]{\color{Magenta}#1 \color{Black}}
\newcommand{\royal}[1]{\color{Royal}#1 \color{Black}}
\newcommand{\orange}[1]{\color{Orange}#1 \color{Black}}
\newcommand{\purple}[1]{\color{Purple}#1 \color{Black}}
\newcommand{\mahogany}[1]{\color{Mahogany}#1 \color{Black}}
\newcommand{\yorange}[1]{\color{YOrange}#1 \color{Black}}
\newcommand{\yellow}[1]{\color{Yellow}#1 \color{Black}}
\newcommand{\rred}[1]{\color{Red}#1\color{Black}}
\newcommand{\bblue}[1]{\color{Blue}#1\color{Black}}
\newcommand{\llblue}[1]{\color{Lblue}#1\color{Black}}
\newcommand{\ggreen}[1]{\color{Green}#1\color{Black}}
\newcommand{\mmagenta}[1]{\color{Magenta}#1\color{Black}}
\newcommand{\rroyal}[1]{\color{Royal}#1\color{Black}}
\newcommand{\oorange}[1]{\color{Orange}#1\color{Black}}
\newcommand{\ppurple}[1]{\color{Purple}#1\color{Black}}
\newcommand{\yyorange}[1]{\color{YOrange}#1\color{Black}}
\newcommand{\yyellow}[1]{\color{Yellow}#1\color{Black}}
\newcommand{\mmahogany}[1]{\color{Mahogany}#1\color{Black}}

\definecolor{GreenYellow}{named}{GreenYellow}
\newcommand{\greenyellow}[1]{\color{GreenYellow}#1\color{Black}}
\definecolor{SkyBlue}{named}{SkyBlue}
\newcommand{\skyblue}[1]{\color{SkyBlue}#1\color{Black}}
\definecolor{Apricot}{named}{Apricot}
\newcommand{\apricot}[1]{\color{Apricot}#1\color{Black}}


\section{Introduction}

There is a world-wide consensus that the
International Linear $e^+e^-$ Collider (ILC) is the next major project
in the field of high-energy physics~\cite{consensus}.
This has most recently been confirmed by the EPP2010
report~\cite{epp2010}. The decision about the start of construction of
the ILC will be taken by the end of this decade. It could go into
operation in the second half of the next decade. It will therefore
take data several years after the start of the Large Hadron Collider
(LHC). The physics case for the ILC, independent of what the LHC will
find, has been made in various studies in the past
years~\cite{tdr,nlc,jlc,lhcilc,Snowmass05Higgs}. The complementarity
and the synergy of the two colliders and combined physics analyses has
been discussed extensively in \citere{lhcilc}.

A very important consideration in respect of the ILC physics case is
the question what the ILC can add to the LHC (or what can a combined
ILC/LHC analysis add to the LHC). It has been
shown~\cite{tdr,nlc,jlc,lhcilc,Snowmass05Higgs} that in {\em all}
conceivable physics scenarios the ILC can add valuable and important
information. In particular, it can add precision analyses,
pinning down model parameters extremely precisely. Moreover, in
contrast to the LHC, this can often be done in a model-independent
way. Furthermore in many scenarios the ILC can discover new states that
cannot be detected at the LHC. The combination of these three
capabilities (precision measurements, model independent analyses,
discovery of new particles) enables the ILC to determine the
underlying physics model. 

While the physics case for the ILC has been made and the physics
potential of the ILC has been analyzed, there are still many tasks
that have to be performed until the full potential of the ILC can be
exhausted. This concerns the experimental analyses as well as the
(corresponding) theoretical calculations. In many scenarios the
feasibility of the experimental analyses has to be worked out in
detail. Theory calculations at the level of the anticipated ILC
precision still have to be performed (see e.g.\
\citeres{DidiLoops,lcwsParisLoops,selfcite}).  
Progress in both directions has to be made over the next years in
order to be ready once the ILC operation starts. A status of the field
and about recent progress was given at the {\em LCWS06} in Bangalore,
India. Here we briefly review the presentations about new experimental
analyses and new theory calculations given at the {\em LCWS06}. 
We focus on the fields of Higgs physics, top/QCD and loop
calculations. The overview about the other fields is given
elsewhere~\cite{physicssummary2}.


\section{Higgs physics}
\label{sec:higgs}

If a (SM-like) Higgs mechanism is realized in nature, 
the LHC will find a Higgs boson and measure it
characteristics~\cite{atlastdr,cmshiggs,HcoupLHCSM,schumi}. 
To be certain the state observed is indeed the Higgs boson, it is 
necessary to measure the couplings of this state to
the $W$~and $Z$~gauge bosons, and to fermions such as the top and
bottom quarks and the tau leptons. 
Consequently, the measurements at the LHC
include a mass determination at the per-cent level and coupling
constant determination at the level of 
10-20\%. However, in order to do this several assumptions about the
realization of 
the Higgs mechanism have to be made. Analyses could become much more
involved if the Higgs boson decay rates are strongly
different from the SM rates. Interesting
physics could easily hide in the 10-20\% precision achievable for the
Higgs boson couplings. Higgs self-couplings are extremely
complicated if not impossible to measure at the LHC. On the other
hand, all these problems can be overcome with the ILC
measurements~\cite{tdr,nlc,jlc,Snowmass05Higgs}.

The progress that will be necessary to fully exploit the ILC
capabilities has been summarized in \citere{Snowmass05Higgs}, where
the main open issues are:
\begin{itemize}
\item 
analyses with full simulations of the relevant ILC processes have to
be performed, 
\item
higher precision in the theory calculations for the relevant processes
are needed to match the anticipated ILC accuracy, 
\item
analyses for SM-like Higgs bosons with ``larger'' mass 
($\MH \gsim 150 \gev$) have to be done, 
\item
tools (encoding high-precision calculations) have to become
available, 
\item
the LHC/ILC interplay has to be worked out in more detail. 
\end{itemize}
These issues have (partially) been addressed at the {\em LCWS06}. 
Progress has been reported e.g.\ about the following subjects (more details
can be found in the original publications):
\begin{itemize}
\item
Improvements in the experimental analyses of triple Higgs boson
couplings~\cite{hhh}. 
\item
Theory calculations for Higgs production at the $\ga\ga$ collider in
the Minimal Supersymmetric Standard Model (MSSM) at the one-loop
level~\cite{gagaHA}. 
\item
Higgs production in models with universal extra
dimensions~\cite{UEDHiggs}. 
\item
Experimental analysis for the determination of anomalous Higgs
couplings at the $e\ga$ collider~\cite{anomHiggs}. 
\item
Doubly charged Higgs production at the $e^-e^-$
collider~\cite{ccHiggs}. 
\item
Measurement of $\tb$ in the MSSM via heavy Higgs boson
production~\cite{tbHA}. 
\item
Progress for the MSSM Higgs tool {\tt FeynHiggs}~\cite{feynhiggs}. 
\end{itemize}


\section{Top/QCD}
\label{sec:topqcd}

Top and QCD physics are a guaranteed physics case for the ILC.
The top quark is deeply connected to many other issues of high-energy
physics: 
\begin{itemize}
\item
Is it just a heavy quark, or does it play a special role in/for
electroweak symmetry breaking?
\item
The experimental uncertainty of $\mt$ induces the largest parametric
uncertainty in the calculation of electroweak precision
observables~\cite{deltamt} and can thus obscure new physics effects.
\item
In supersymmetric (SUSY) models the top quark mass is an important
input parameter and drives spontaneous symmetry breaking and
unification. 
\item
Little Higgs models contain ``heavier tops''.
\end{itemize}
The status of the field has been summarized in \citere{Snowmass05top}. 
Especially the calculations for $e^+e^- \to t \bar t$ at the 
threshold are quite advanced, see e.g.\ \citere{ttcalc} for a review or
\citere{ttcalc2} for a recent update. Also for the process
$e^+e^- \to t \bar t H$ and the determination of the top Yukawa coupling 
substantial progress has been made recently, see 
e.g.\,\citeres{ttHfarrell,ttHjuste}.

Advance in the field of top/QCD has been reported at the {\em LCWS06} in the
following subjects (more details can again be found in the original
publications): 
\begin{itemize}
\item
Improved calculation of the cross section 
$\si(\ga\ga \to \mbox{hadrons})$~\cite{gghad}.
\item
A new theory evaluation on how to use the lepton characteristics in
top decays as probe of new physics scenarios~\cite{topdecay}.
\item
A calculation of $\ga\ga \to H,A \to t \bar t$ in the MSSM, including
the polarization of the top quarks; this measurement can be used to
test the mixing between $\cp$-even and $\cp$-odd Higgs
bosons~\cite{HAmix,HAmix2}. 
\end{itemize}

In order to exploit the full ILC capabilities for top quark physics,
many improvements are still necessary. This includes 
a coherent treatment of electroweak effects at the 
$t\bar t$~threshold~\cite{Snowmass05top} 
or the development of tools
providing the prediction of $e^+e^- \to 6f$ or $8f$ (see e.g.\
\citere{lusifer}). 


\section{Loop Calculations}

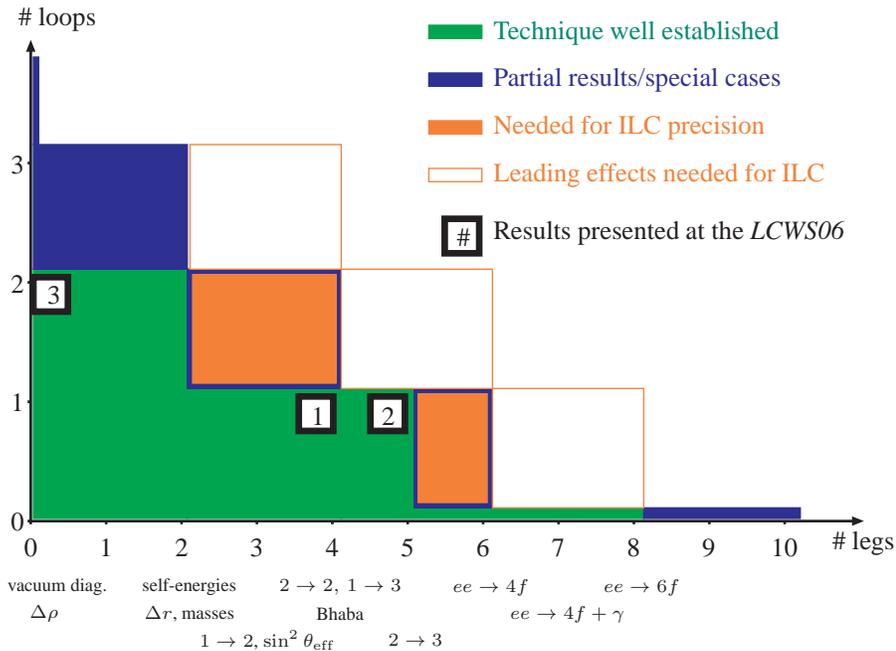
\begin{figure}[htb!]
\begin{center}
\setlength{\unitlength}{0.5pt}
\begin{picture}(650,400)
  \LinAxis(0,0)(312.5,0)(10.999,1,-1,0,1)
  \LongArrow(0,0)(312.5,0)
  \LinAxis(0,0)(0,180)(3.999,1,+1,0,1)
  \LongArrow(0,0)(0,180)
  \Text(-10,  0)[c]{0}
  \Text(-10,090)[c]{1}
  \Text(-10,180)[c]{2}
  \Text(-10,270)[c]{3}
  \Text(20,380)[c]{\# loops}
  \Text(  0,-20)[c]{0}
  \Text( 57,-20)[c]{1}
  \Text(114,-20)[c]{2}
  \Text(171,-20)[c]{3}
  \Text(228,-20)[c]{4}
  \Text(285,-20)[c]{5}
  \Text(342,-20)[c]{6}
  \Text(399,-20)[c]{7}
  \Text(456,-20)[c]{8}
  \Text(513,-20)[c]{9}
  \Text(570,-20)[c]{10}
  \Text(630,-17)[c]{\# legs}
\Text( 20,-50)[c]{\scriptsize vacuum diag.}
\Text( 10,-70)[c]{\scriptsize $\De\rho$}
\Text(120,-50)[c]{\scriptsize self-energies}
\Text(120,-70)[c]{\scriptsize $\De r$, masses}
\Text(179,-90)[c]{\scriptsize $1 \to 2$, $\sweff$}
\Text(234,-50)[c]{\scriptsize $2 \to 2,\, 1 \to 3$}
\Text(234,-70)[c]{\scriptsize Bhaba}
\Text(291,-90)[c]{\scriptsize $2 \to 3$}
\Text(348,-50)[c]{\scriptsize $ee \to 4f$}
\Text(406,-70)[c]{\scriptsize $ee \to 4f + \ga$}
\Text(462,-50)[c]{\scriptsize $ee \to 6f$}

\CBox(  1,  1)( 59, 95){Green}{Green}
\CBox( 59,  1)(117, 50){Green}{Green}
\CBox(117,  1)(231,  5){Green}{Green}
\CBox(  1, 95)( 59,142){Blue}{Blue}
\CBox(  1,142)(  3,175){Blue}{Blue}
\CBox( 59, 50)(117, 95){Blue}{Blue}
\CBox(231,  1)(290,  5){Blue}{Blue}
\CBox(144,  5)(174, 50){Blue}{Blue}
\CBox( 61, 52)(115, 93){Orange}{Orange}
\CBox(117,  5)(144, 50){Green}{Green}
\CBox(146,  7)(172, 48){Orange}{Orange}
\CBox( 60, 95)(117,142){Orange}{White}
\CBox(117, 50)(174, 95){Orange}{White}
\CBox(174,  5)(231, 50){Orange}{White}

\CBox(150,182)(170,187){Green}{Green}
\Text(350,370)[l]{\green{Technique well established}}
\CBox(150,164)(170,169){Blue}{Blue}
\Text(350,334)[l]{\blue{Partial results/special cases}}
\CBox(150,146)(170,151){Orange}{Orange}
\Text(350,298)[l]{\orange{Needed for ILC precision}}
\CBox(150,128)(170,133){Orange}{White}
\Text(350,262)[l]{\orange{Leading effects needed for ILC}}
\CBox(155,100)(170,115){Black}{Black}
\CBox(157,102)(168,113){Black}{White}
\Text(330,215)[c]{\black{\#}}
\Text(350,218)[l]{\black{Results presented at the {\em LCWS06}}}

\CBox(100, 33)(115, 48){Black}{Black}
\CBox(102, 35)(113, 46){Black}{White}
\Text(220, 80)[c]{\black{1}}
\CBox(127, 33)(142, 48){Black}{Black}
\CBox(129, 35)(140, 46){Black}{White}
\Text(273, 80)[c]{\black{2}}
\CBox(000, 78)(015, 93){Black}{Black}
\CBox(002, 80)(013, 91){Black}{White}
\Text(020,171)[c]{\black{3}}

\end{picture}
\end{center}
\vspace{1.5cm}
\caption{
Status of the field of loop calculations.
The light shaded (orange) area shows
what will be needed to match the anticipated ILC precision.
The squares and numbers indicate the contributions presented at the
{\em LCWS06}. 
}
\label{fig:loops}
\end{figure}

The ILC will provide measurements of masses,
couplings and cross sections at (or even below) the per-cent
level~\cite{tdr,nlc,jlc}. A special example is the GigaZ option, which will
determine the $W$~boson mass with an error of $7 \mev$ and the
effective leptonic weak mixing angle with a precision of 
$1.3 \times 10^{-5}$, see \citeres{GigaZexp,blueband} and references
therein. These anticipated future  
high precision measurements can only fully be exploited if they are matched
with a theory prediction at the same level of accuracy. 
These theory predictions have to be obtained in the model under
investigation (e.g.\ the SM, or the MSSM,
for a recent overview see \citere{PomssmRep}). 
The status of the field of loop calculations (including the
presentations at the {\em LCWS06}) is briefly summarized in
\reffi{fig:loops}%
\footnote{
We are grateful to S.~Dittmaier for helpful discussions.
}%
.~The complication of a higher-order loop calculation increases with
the number of loops as well as with the number of external legs. On
the other hand, it also increases with the number of (mass) scales
appearing in
the loop integral. While one-scale integrals (as occur e.g.\ in QCD)
are usually the easiest possibility for a certain loop topology, two
or more scales make the evaluation increasingly difficult. This poses
a special problem in SUSY, where many independent mass
scales can appear in one loop diagram. 
In \reffi{fig:loops} on the horizontal axis the
number of legs, and on the vertical one the number of loops is
shown. Accordingly, the number of scales has to be kept in mind for
the individual contributions reviewed below.

The medium shaded (green) area in \reffi{fig:loops} displays the
number of loops and legs for which the techniques are meanwhile well
established, even for an arbitrary number of scales. 
For these cases often public algebraic computer codes exist that
do the main part of the loop calculation itself, for an overview see
e.g.\ \citere{DidiLoops}.
The dark shaded (blue) area corresponds to the number of loops and legs
for which partial results or calculation for special cases have been
performed. This represents today's frontier of the field of loop
calculations. 

Compared to the situation about two years
ago~\cite{lcwsParisLoops} significant progress has been made.
Three examples for recent progress in the field of loop calculations
are 
\begin{itemize}
\item
a complete \order{\al} calculation for $e^+e^- \to 4f$~\cite{ee4f}, 
\item
a complete \order{\al} calculation for 
$e^+e^- \to \nu \bar\nu HH$~\cite{eennHH}, 
\item
the automation of $2 \to 3$ processes at the one-loop level in the SM
and the MSSM (including now also NMFV effects)~\cite{FANMFV}. 
\end{itemize}

In order to match the anticipated experimental precision of a future
ILC, the field of loop calculations still has to advance substantially.
The necessary improvement is indicated in \reffi{fig:loops} as the
light shaded (orange) areas, which will have to be under full control
for the ILC precision%
\footnote{
Progress is also needed to improve {\em current} electroweak precision
analyses, as reported in other sessions at the {\em LCWS06}, see e.g.\
\citeres{lcws06ehow,MWweber}. 
}%
. 

Some advance has been presented at this conference, which
is shown as black rectangles (and numbers):
\begin{itemize}
\item[\#1]
The GRACE system has been extended to $2 \to 2$ processes at the
one-loop level in the MSSM~\cite{GraceMSSM}. 
\item[\#2]
Improvements in the calculation of $Q_T$ spectra for $2 \to 3$
processes at the one-loop level have been obtained~\cite{berger06}. 
\item[\#3]
Two-loop vacuum diagrams in the MSSM with complex phases have been
evaluated~\cite{mhcMSSM2L}. 
\end{itemize}
By comparing the necessary level of loop calculations and the current status
(including the progress reported at the {\em LCWS06}), it becomes aparent that
the field of loop calculations deserves a lot of attention in the next years. 


\section{Conclusions}

There are still many tasks
that have to be performed until the full potential of the ILC can be
exploited. 
We reviewed the
status of the field and the recent progress that was reported at the
{\em LCWS06} in Bangalore (India) in the fields of Higgs physics,
top/QCD and loop calculations. While progress has been achieved, 
still continuous progress over the next years will be necessary in the 
experimental analyses as well as the
(corresponding) theoretical calculations in 
order to be ready once the ILC operation starts.


%



\end{document}